\documentclass[10pt,aps,prd,twocolumn,showpacs,amsmath,amssymb,nofootinbib,eqsecnum,preprintnumbers,superscriptaddress]{revtex4-2}
\usepackage{amsmath,amssymb}
\usepackage{enumitem}  
\usepackage[usenames, dvipsnames]{color} 
\usepackage{graphicx} 
\usepackage{comment}
\usepackage[normalem]{ulem}
\usepackage[utf8]{inputenc}
\usepackage{url}
 \usepackage{xcolor}

\newcommand{\dd}{{\rm d} }
\usepackage{hyperref}

\usepackage{graphicx}% Include figure files
\usepackage{dcolumn}% Align table columns on decimal point
\usepackage{bm}% bold math
\newcommand{\be}{\begin{equation}}
\newcommand{\ee}{\end{equation}}
\newcommand{\ba}{\begin{eqnarray}}
\newcommand{\ea}{\end{eqnarray}}

\newcommand{\beq}{\begin{equation}}
\newcommand{\eeq}{\end{equation}}
\newcommand{\beqa}{\begin{eqnarray}}
\newcommand{\eeqa}{\end{eqnarray}}

\begin{document}

\title{Excising Cauchy Horizons with Nonlinear Electrodynamics}
\author{Tom\'a\v{s} Hale}
\email{tomas.hale@mff.utf.cuni.cz}
\affiliation{Institute of Theoretical Physics Faculty of Mathematics and Physics,
Charles University V Holešovičkách 2, 180 00 Prague 8, Czech Republic}

\author{Robie A. Hennigar}
\email{robie.a.hennigar@durham.ac.uk}
\affiliation{Centre for Particle Theory, Department of Mathematical Sciences,
\\
Durham University, Durham DH1 3LE, U.K.}

\author{David Kubiz\v{n}\'ak}
\email{david.kubiznak@matfyz.cuni.cz}
\affiliation{Institute of Theoretical Physics Faculty of Mathematics and Physics,
Charles University V Holešovičkách 2, 180 00 Prague 8, Czech Republic}

%\date{\today}
\date{June 25, 2025}

\begin{abstract}

Charged and/or rotating black holes in General Relativity feature Cauchy horizons, which indicate a breakdown of predictability in the theory.  Focusing on spherically symmetric charged black holes, we remark that the inevitability of Reissner--Nordstr{\"o}m Cauchy horizon is due to the divergent electromagnetic self-energy of point charges. We demonstrate that any causal theory of nonlinear electrodynamics that regularizes the point charge self-energy also eliminates Cauchy horizons for weakly charged black holes. These black holes feature one (event) horizon and a spacelike singularity, analogous to the Schwarzschild metric. An example with Born--Infeld electrodynamics illustrates how this gives rise to an upper bound on the charge, which we compare with known bounds.

\end{abstract}

\maketitle

\section{Introduction}
\label{sc:intro}

Among all the effects that General Relativity (GR) provides explanations for, {\em black holes} describe the most radical physical scenario, where an entire region becomes causally separated from the rest of the universe. The nature of causality in GR becomes ever more obscure and complex upon considering the maximal 
{\em analytical extension}
of various black hole spacetimes. Major issues occur already when inspecting the extended spacetime causal structure in the simple case of adding an electric charge to the static spherically symmetric %(SSS) 
black hole, described by the Reissner--Nordstr\"om solution, or even the most astrophysically relevant case: the rotating black hole, described by the Kerr geometry.

It seems that GR has the grace of prohibiting us from seeing any pathologies that lack any further explanation within the theory, such as curvature singularities, closed timelike curves, and Cauchy horizons, hiding them in these special regions \cite{Penrose:1969pc}. However, the very existence of Cauchy horizons is problematic, as they signal the theory has lost its predictive power. It is widely believed that Cauchy horizons in GR are an artefact of \textit{fine tuning}. This is because inner horizons are susceptible to both classical mass inflation ~\cite{Poisson:1989zz} and (even stronger) semi-classical instabilities~\cite{Hollands:2019whz}. As a result, a singularity likely forms, giving rise to a {\em modified black hole spacetime} with a single horizon and spacelike singularity {\`a} la Schwarzschild. It therefore seems interesting to seek (singular) black hole solutions that, despite featuring charges and/or rotation, have no Cauchy horizons (see \cite{Leon:2024uao} for a specific off-shell model for such metrics).

At the same time, there are instances in which some authors would like to avoid a singularity at the inner horizon and preserve in some way the structures that lie beyond it. Such a desire can remain consistent with predictivity so long as the inner horizon is not a Cauchy horizon. This perspective is common, e.g., in the context of the regular black hole program~\cite{Ayon-Beato:2000mjt, Bueno:2024dgm, Carballo-Rubio:2025fnc}.  However, in this case the instabilities which ``save the day'' in GR take on a more problematic role. Mechanisms to resolve the mass inflation instability of inner horizons have been put forward, such as requiring that the inner horizon is degenerate/extremal~\cite{Carballo-Rubio:2022kad}. However, the generality of such a mechanism is questionable, as it requires seemingly ad hoc fine-tuning of various parameters. Moreover, there are indications that while inner extremality is sufficient to avoid the classical instability, it is \textit{not} sufficient to avoid the semi-classical one~\cite{McMaken:2023uue}. Thus, the spacelike curvature singularity such as that of Schwarzschild may in fact be a desirable feature of {\em classical} black hole solutions, at least when we consider dynamics driven by GR.

Here we focus on charged, spherically symmetric black holes and investigate the sensitivity of the Cauchy horizon to \textit{deformations} of Maxwell's theory. We emphasize the reason the Cauchy horizon in the Reissner--Nordstr{\"o}m spacetime cannot be avoided is because the electromagnetic self-energy in Maxwell theory is necessarily much larger (actually infinite) than the black hole mass. Hence, we restrict our attention to theories of {\em nonlinear electrodynamics (NLE)}, for which one of the prime motivations is their ability to \textit{regularize} the self-energy of point sources~\cite{Born:1934gh, Sorokin:2021tge}. Heuristically, it is easy to understand why NLE would play an important role in the causal structure of charged black holes. For weakly charged black holes, the field strength at the inner horizon can become very large, pushing into the regime where NLE effects dominate.
It is the purpose of our paper to uncover the implications of NLEs in this regime.

Our main result is that in any theory of NLE that satisfies the strong energy condition and has finite self-energy, the corresponding \textit{weakly charged} black hole solutions have precisely the same causal structure as Schwarzschild, i.e.~an event horizon and spacelike curvature singularity. As all four energy conditions are necessary conditions for causality in any NLE, our results encompass all cases of physical interest~\cite{Russo:2024xnh}. More specifically, Schwarzschild-like causal structure is realized whenever the gravitational mass exceeds the electromagnetic self-energy of the charged matter. The case of weakly charged black holes is the most physically relevant, as the accumulation of charge is severely limited by both astrophysical and theoretical mechanisms~\cite{Wald:1974np, Gibbons:1975kk, Levin:2018mzg, Zajacek:2019kla}.

To give a concrete example, we analyze the charged black hole solutions of 
Einstein--Born--Infeld theory  
%\tcr{Born--Infeld theory}
\cite{Born:1934gh}, and show that the absence of the Cauchy horizon imposes an upper bound on the charge of a black hole. Given the known experimental and theoretical constraints on Born--Infeld theory, we discuss how this bound compares with theoretical and astrophysical limits.

\section{Single Horizon Charged Black Holes}

NLE is a framework generalizing Maxwell's classical electromagnetism by allowing the Lagrangian ${\cal L}$ to be a function of the two electromagnetic invariants: 
\be 
{\cal L}={\cal L}({\cal S}, {\cal P}^2)\,,\quad  \mathcal{S} = \frac{1}{2} F_{\mu\nu} F^{\mu\nu}\,,\quad  \mathcal{P} = \frac{1}{2} F_{\mu\nu} \left(*F^{\mu\nu}\right)\,.
\ee 
Here, $F_{\mu\nu}=(dA)_{\mu\nu}$ is the electromagnetic field strength, written in terms of the vector potential $A_\mu$,  {and the Maxwell theory is recovered upon setting ${\cal L}=-\frac{1}{2}{\cal S}$.} 

When minimally coupled to GR, the corresponding generalized Einstein--Maxwell equations take the form: 
\ba 
\dd F&=&0\,,\quad \dd *D =0\,,\quad G_{\mu\nu}=8\pi T_{\mu\nu}\,,\nonumber\\
D_{\mu\nu}&\equiv&-2\frac{\partial {\cal L}}{\partial F^{\mu\nu}}=-2\bigl({\cal L}_{\cal S}F_{\mu\nu}+{\cal L}_{\cal P}(*F)_{\mu\nu}\bigr)\,,\\
T^{\mu\nu}&=&-\frac{1}{4\pi}\Bigl(2 F^{\mu\sigma}{F}^\nu{}_\sigma\mathcal{L}_{\mathcal{S}}+{\cal P}{\cal L}_{\cal P}g^{\mu\nu}-\mathcal{L}g^{\mu\nu}\Bigr)\,,\nonumber
\ea 
where ${\cal L}_{\cal S}=\frac{\partial {\cal L}}{\partial {\cal S}}$ and ${\cal L}_{\cal P}=\frac{\partial {\cal L}}{\partial {\cal P}}$.

Consider a self-gravitating electrically charged spherically symmetric solution. In spherical symmetry, ${\cal P}$ identically vanishes and we have the restricted theory $\mathcal{L}=\mathcal{L}(\mathcal{S})$. Moreover, for any such theory the metric is necessarily described by a single function $f=f(r)$, e.g. \cite{Jacobson:2007tj}, and can be written as
\begin{align}
    \dd s^2 = -f\dd t^2+\frac{\dd r^2}{f}+r^2\dd\Omega^2\,,\quad A =-\phi\dd t\,,
\end{align} 
where $\phi=\phi(r)$, and $\dd\Omega^2=\dd\theta^2+\sin^2\!\theta\dd\varphi^2$ is the standard area element on the sphere. % and the electromagnetic tensor  
The field equations then yield the following solution:
\ba \label{eq:solution}
-2\mathcal{L}_\mathcal{S}E&=&\frac{Q}{r^2}\,, \quad
f=1-\frac{2M}{r}+\frac{q^2(r)}{r^2}\,,\nonumber\\ 
q^2(\rho)&\equiv & -2\rho\int_\rho^\infty r^2(2{\cal L}_{\cal S}E^2+{\cal L})\,\dd r\,, \label{eq:reduced EE}
\ea
where $E=-\phi'(r)$. Here, $Q$ is an integration constant. It is related to the asymptotic electric charge by assuming that the field strength approaches that of Maxwell in the weak field limit (at large distances), i.e. ${\cal L}_{\cal S}\to -\frac{1}{2}$. Since ${\cal S}=-E^2$, the first equation represents an algebraic equation for $E$. Similarly, $M$ is the `gravitational mass' integration constant and we have  
$\lim_{r\to \infty}q(r)= Q$.

An important observation in what follows is the connection between the function $q^2(r)$ appearing in the metric and the self-energy of a point particle in the specified theory of NLE. Explicitly, we define
\be 
U_{\mbox{\tiny self}}(r)=\frac{q^2(r)}{2r}=- \frac{1}{4 \pi} \int T^t_t\, {\rm d} V \,,\label{eq:self energy}
\ee 
which represents the electrostatic self-energy stored in a region between the radius $r$ and infinity. The self-energy of a point source given by $\lim_{r\to 0} U_{\mbox{\tiny self}}(r)$~\cite{Hoffmann:1935ty, Hennigar:2025ftm}. 

In the case of Maxwell theory, the above evaluates to $U_{\rm self} = Q^2/(2r)$. At short distances, the energy stored between $r$ and infinity becomes large, blowing up in the strict $r\to 0$ limit. Hence, at sufficiently short distances the matter terms \textit{necessarily} overwhelm the $2M/r$ gravitational potential, driving the metric function to $f(r\to 0) \to +\infty$. It is this short distance behaviour of the self-energy that ultimately ensures that the Cauchy horizon is compulsory.

Let us now assume the following expansion of $U_{\mbox{\tiny self}}$ near the origin $r=0$: 
\be 
U_{\mbox{\tiny self}}=\frac{U_{\mbox{\tiny self}}^{(-1)}}{r}+U_{\mbox{\tiny self}}^{(0)}+r U_{\mbox{\tiny self}}^{(1)}+O(r^2) \,.
\ee
The absence of higher inverse powers of $r$ indicates that we are assuming the NLE is not \textit{more divergent} than Maxwell theory, for which the series contains only one nonzero term, $U_{\mbox{\tiny self}}^{(-1)}=Q^2/2$. In terms of the above we have at small $r$
\be
f=\frac{2 U_{\mbox{\tiny self}}^{(-1)}}{r^2}+\frac{2(U_{\mbox{\tiny self}}^{(0)}-M)}{r}+1+2 U_{\mbox{\tiny self}}^{(1)}+O(r)\,.
\ee
In any theory that, like Maxwell's theory, has a non-zero (and positive) $U_{\mbox{\tiny self}}^{(-1)}$, Cauchy horizons will be unavoidable. In any such theory, adding a Coulomb charge to a Schwarzschild black hole cannot be viewed as a perturbation -- the norm of the difference in $f$ before and after adding such a charge is in no way small.

To proceed, let us then assume a finite self-energy, i.e. $U_{\mbox{\tiny self}}^{(-1)}=0$ and  $0<U_{\mbox{\tiny self}}^{(0)}<\infty$, and examine the possible behaviours for the black hole interior. The dominant behaviour at small radius is determined by the ratio of the self-energy $U_{\mbox{\tiny self}}^{(0)}$ and gravitational mass $M$. If the self energy is smaller than the gravitational mass, $U_{\mbox{\tiny self}}^{(0)}<M$, then the metric function $f\to -\infty$ and we have a spacelike singularity. Moreover, provided that the NLE satisfies the strong energy condition we can further deduce that there is generically \textit{exactly one} horizon in this case --- see Appendix. Thus, this branch of solutions is free of Cauchy horizons;  as per usual we call the corresponding branch of black holes the {\em Schwarzschild-like (S)-branch}.

On the other hand, when the gravitational mass is smaller than the self energy, $M<U_{\mbox{\tiny self}}^{(0)}$, the central singularity is timelike. Provided that the NLE satisfies the strong energy condition, we can be assured that there is \textit{at most} one inner horizon in this case. Thus, the causal structure in this case is identical to Reissner--Nordstr\"om --- we have either an event and inner horizon, or a naked timelike singularity.  We call the corresponding branch the {\em Reissner--Nordstr\"om-like (RN)-branch.}

Finally, the case when $M=U_{\mbox{\tiny self}}^{(0)}$ is {\em marginal} and corresponds to finite $f$ at the origin. Despite a regular metric function, there is still a curvature singularity at $r = 0$. Whether this singularity is spacelike, timelike, or null depends on the subleading terms in the expansion of $U_{\mbox{\tiny self}}$. (To obtain a regular black hole spacetime, one would have to further demand that $U_{\mbox{\tiny self}}^{(1)}=0$ in addition to $M=U_{\mbox{\tiny self}}^{(0)}$ to obtain the necessary Minkowksi/(A)dS core; see however a no-go theorem in \cite{Bronnikov:2000vy}.)

In all cases, when $Q \to 0$ we have $U_{\mbox{\tiny self}}^{(0)} \to 0$ and the Schwarzschild metric is recovered. The weakly charged black holes all belong to the S-branch. Hence, these are charged black holes for which the causal and horizon structure is Schwarzschild-like. This works precisely because the self-energy no longer blows up near the origin.

\section{An example: Born--Infeld electrodynamics}

Let us now illustrate the discussion of the previous section for the most well-known NLE model, Born--Infeld theory \cite{Born:1934gh}. Born--Infeld theory enjoys a number of desirable properties. The theory predicts a finite self-energy for point sources, is causal, birefringence-free, respects electromagnetic duality even at the nonlinear level, and has a stress tensor which satisfies the dominant and strong energy conditions~\cite{Born:1934gh,Plebanski:1970zz,Bialynicki-Birula:1992rcm,Gibbons:1995cv,  Russo:2024xnh}. On the other hand, Born-Infeld theory  also arises naturally as an effective low energy theory for gauge fields in string theory~\cite{Fradkin:1985qd}  and in D-brane physics \cite{Leigh:1989jq}. 

The Lagrangian of Born--Infeld electrodynamics reads~\cite{Born:1934gh}
\begin{align}
\mathcal{L}_{\text{BI}}=b^2\bigg(1-\sqrt{1+\frac{\mathcal{S}}{b^2} - \frac{\mathcal{P}^2}{4 b^4}}\bigg)\,,
\end{align}
where $b$ represents an upper bound on the field strength, with the Maxwell theory recovered in the limit $b\to \infty$. The corresponding self-gravitating solution reads
\ba 
f&=&1-\frac{2M}{r}+\frac{2b^2}{r}\int_r^\infty \!\bigg(\sqrt{r^4+\frac{Q^2}{b^2}}-r^2\bigg) {d} r\,, \nonumber\\
E&=&\frac{Q}{\sqrt{r^4+Q^2/b^2}}\,.
\ea 
%Note that the NLE parameter $b$ yields the the maximum field strength of $E$ in the origin. 
The electromagnetic self-energy in this case is finite and reads  
\begin{align} 
U_{\mbox{\tiny self}}&=b^2\!\!\int_r^\infty\!\!\! \Big(\sqrt{r^4+\frac{Q^2}{b^2}}-r^2\Big) { d}r
=U_{\mbox{\tiny self}}^{(0)}-bQr+O(r^2)\,,\ 
\nonumber\\
U_{\mbox{\tiny self}}^{(0)}&=\frac{1}{6}\sqrt{\frac{b}{\pi}}|Q|^{3/2}\Gamma\Bigl(\frac{1}{4}\Bigr)^2\,.
\end{align}
\begin{figure*}[t]
\begin{center}
\includegraphics[width=0.45\textwidth]{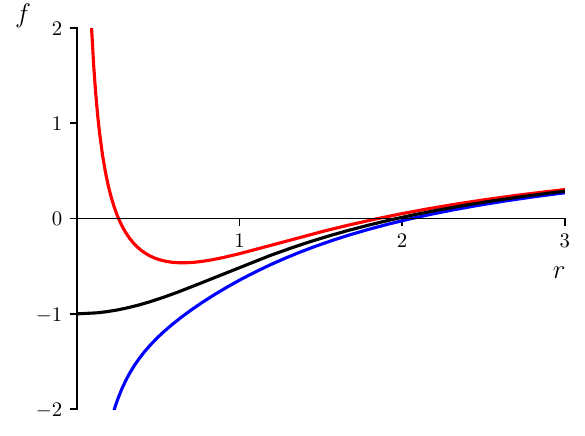}
\quad \includegraphics[width=0.45\textwidth]{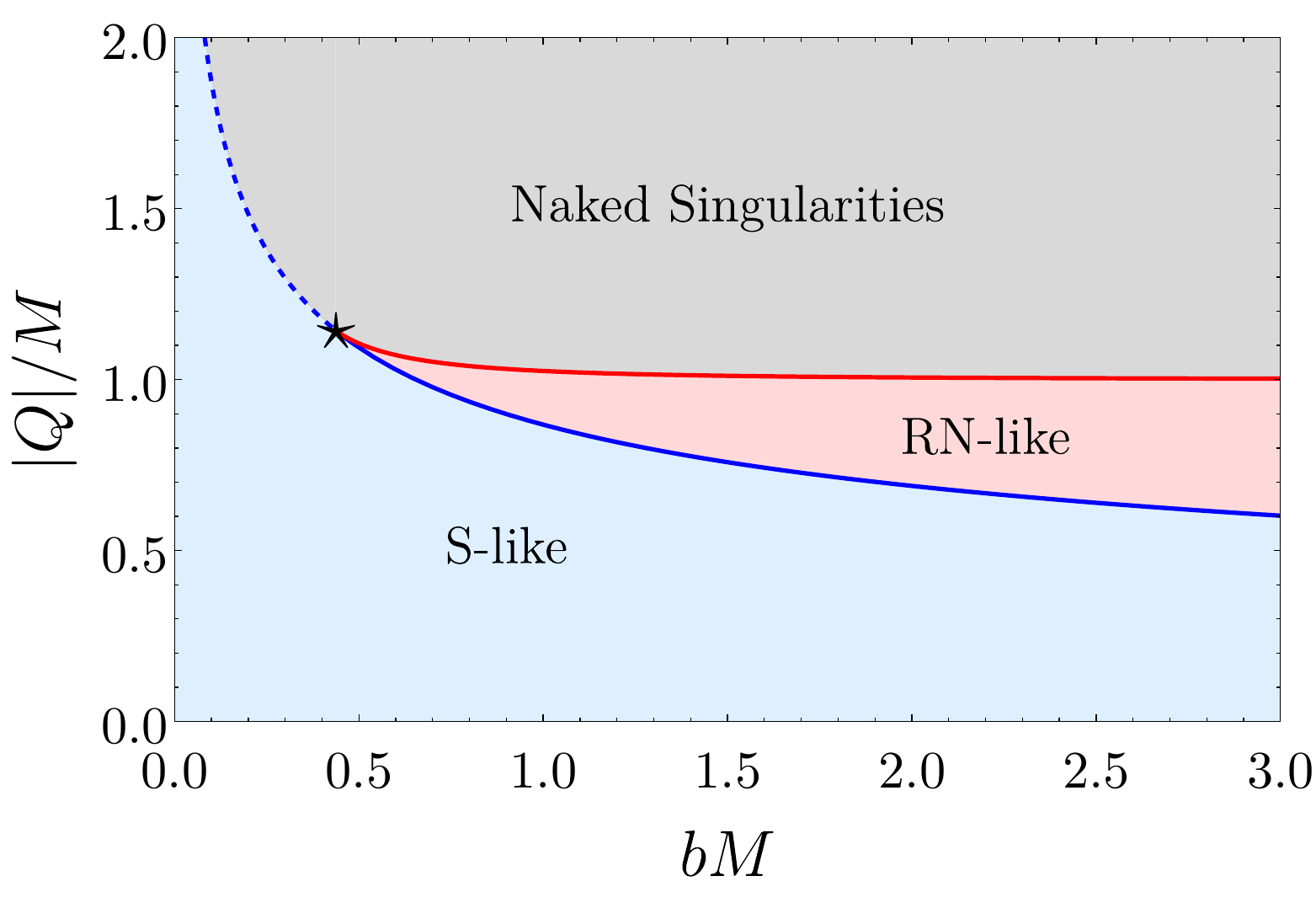}
\caption{ {\bf Born--Infeld black holes.} {\it Left}: Three distinct behaviours are possible for charged black holes in BI electrodynamics depending on the parameters $M,Q,b$: the S-branch occurs when $M>U_{\mbox{\tiny self}}^{(0)}$; it is represented by the bottom blue curved and features a spacelike singularity and single horizon. The RN-branch, for $M<U_{\mbox{\tiny self}}^{(0)}$, shown by the red upper curve, possesses a timelike singularity and an unwanted feature of the Cauchy horizon; the marginal case when the black hole mass equals the electrostatic energy $M=U_{\mbox{\tiny self}}^{(0)}$ is displayed by the black middle curve. {\it Right}: 
Parameter space diagram for the possible behaviours of the metric function $f$: the blue-shaded region corresponds to S-like black holes, the red-shaded region to RN-like black holes, and the gray shaded region to naked singularities. The blue line represents the marginal black holes (and coincides with the bound \eqref{eq:bound}). When the line is solid, the marginal solution is a black hole, while the marginal solution is a naked singularity when the line is dashed. The solid red line represents the extremal BI black holes. The star is the point where the marginal and extremal black holes coincide -- this solution has a null singularity.}
\label{Fig:fBI}
\end{center}
\end{figure*} 

In the left panel of Fig. \ref{Fig:fBI} we display the behaviour of the metric function. We observe the S-branch (blue curve) with a single horizon and spacelike singularity for weakly charged black holes and the RN-branch (red curve) with a timelike singularity and the Cauchy horizon (or naked singularity). The right panel of Fig. \ref{Fig:fBI} illustrates the phase space of solutions as a function of the charge-to-mass ratio and the mass at fixed Born--Infeld parameter. The existence of S-type and RN-type black holes in Born--Infeld theory has long been known, e.g.~\cite{deOliveira:1994in, Fernando:2003tz}, however the broader context and generality of this behaviour described here is new. 

Consider the RN-branch of solutions, which always terminate with an extremal black hole as the charge is increased. Extremal black holes in Born--Infeld theory have particular bounds on their charge-to-mass ratio that differ from Maxwell theory,
\be 
1 \le \left(\frac{|Q|}{M}\right)_{\!\rm ext}  \le \frac{6 \sqrt{\pi}}{\Gamma(1/4)^2} \, .
\ee
 The lower bound is achieved in the limit $b M \to \infty$  while the upper bound is exactly when the marginal solution is extremal which requires
\be 
M = M_\star \equiv \frac{\Gamma(1/4)^2}{12 \sqrt{2 \pi} b} \, , \quad |Q| = Q_\star \equiv \frac{1}{2b} \, .
\ee
The RN-branch ceases to exist for $M < M_\star$, for these smaller masses black holes are necessarily S-type. 
In particular, note that S-type Born--Infeld black holes of \textit{arbitrarily large} charge-to-mass ratio can exist in principle, in contrast to the situation in Maxwell theory.

For $M > M_\star$, prohibiting the RN-branch (or the Cauchy horizon) entirely imposes an upper bound on the charge of a black hole derived from the requirement $M>U_{\mbox{\tiny self}}^{(0)}$,
\be \label{eq:bound}
\frac{|Q|}{M}<\Bigg(\frac{6}{\Gamma(1/4)^2}\sqrt{\frac{\pi}{bM}}\Bigg)^{2/3}\, .
\ee 
On the other hand, for $M < M_\star$ this is precisely the bound required to ensure the absence of naked singularities. Hence for all masses, this provides a bound on the charge for which the black holes have physically sensible interiors. At fixed Born--Infeld parameter, this bound is tightest for large mass black holes while it is not particularly restrictive in the limit of very small masses. For a fixed mass, the ratio tends to zero in the Maxwell limit $b \to \infty$, as required.

While cosmic censorship motivates the bound when $M < M_\star$, \textit{requiring} it for $M > M_\star$ is in principle ad hoc. However, as we will now discuss, for physically viable values of the Born-Infeld parameter, the bound is actually very large by astrophysical standards. In their original paper, Born and Infeld estimated the value of $b$ by requiring that the electron's self-energy is equal to its mass~\cite{Born:1934gh}. This estimate remains in line with the best modern constraints, e.g.~\cite{Davila:2013wba}, which require that $b 	\gtrsim 10^{20} \, \, {\rm V/m}$.\footnote{In geometric units, $10^{20} \, \, {\rm V/m} \approx 9.59 \times 10^{-8} \, \, {\rm m}^{-1}$.} Hence, converting to SI units we have 
\be \label{eq:BI-bound}
Q_{\rm max} \approx 2.85  \left(\frac{M}{M_{\odot}} \right)^{2/3} \left(\frac{b_0}{b} \right)^{1/3} \times 10^{21}   \, \, {\rm C} \, .
\ee
Here $M_{\odot}$ is one solar mass, while $b_0 = 10^{20} \, \, {\rm V/m}$ is taken as a reference scale for the Born--Infeld parameter. 

A strict fundamental limit on charge accumulation arises from Schwinger pair production in the vicinity of the event horizon~\cite{Gibbons:1975kk}. This will happen when the horizon electric field is $E \sim m_e^2/(\hbar e)$ in geometric units. This yields an upper bound on the charge $Q_{\rm max}^{\rm pp}  \sim m_e^2 M^2/(\hbar e)$.  If we equate pair production bound with the NLE bound, we find that $Q_{\rm max}^{\rm NLE} > Q_{\rm max}^{\rm pp} $ provided that
\be 
M \lesssim \left(\frac{\hbar^3 e^3}{m_e^6 b} \right)^{\frac{1}{4}}  \sim 10^5 \left(\frac{b_0}{b} \right)^{\frac{1}{4}} \, \, M_{\odot} \, .
\ee
Hence, for all but the largest supermassive black holes, the inner horizon would be eliminated by NLE effects. 

On the other hand, in any realistic environment the charge of black holes is well below the fundamental limit implied by the Schwinger effect. The bound is tightest for large black holes, so let us assess it for Sagittarius A* which has $M \approx 4.3 \times 10^{6} \,  M_{\odot}$. Taking $b = b_0$, we find $Q_{\rm max}^{\rm Sgr \, A^*} \approx 7.55 \times 10^{25} \, \, {\rm C} \, .$ Comparing this with the observational constraints on the charge  $Q_{\rm obs}^{\rm Sgr \, A^*} \lesssim 3 \times 10^{8} \, \, {\rm C}$ and  theoretical upper limit of $Q_{\rm theo}^{\rm Sgr \, A^*}\approx 10^{15} \, \, {\rm C}$~\cite{Zajacek:2018vsj}, we see that both comfortably satisfy the bound. On the other hand, we may take the Planck scale as a firm upper limit to the value of $b$.\footnote{In the stringy realization of Born--Infeld theory we have $b \sim \ell_s^{-1}$ where $\ell_s$ is the string length.
While $\ell_s$ is often assumed to be comparable to the Planck length, it is not a strict requirement.
} If we take $b \sim 1/\ell_P$ where $\ell_P$ is the Planck length, then we find $Q_{\rm max}^{\rm Sgr \, A^*} \approx 8.72 \times 10^{11} \, \, {\rm C}$, still well above the observational limit.

Of course, the main point of these calculations is not to make an experimental prediction, but rather to highlight that the bound is actually physically reasonable. Thus, the absence of Cauchy horizons requires that black holes are weakly charged, but this charge is still large by astrophysical standards.  Hence, the issue of the Reissner--Nordstr{\"o}m Cauchy horizon becomes irrelevant for practical purposes.

While we have focused on the case of Born--Infeld electrodynamics, the main lessons of this analysis are quite general. Qualitatively similar conclusions hold for any other self-energy regularizing NLE, and in fact it is possible to prove on general grounds that the maximum charge will always have precisely the same scaling with mass and field strength as in~\eqref{eq:BI-bound}. To see this, suppose that we consider a NLE for which strong field effects become important at a scale $b$. Then, we can expect deviations from the Reissner-Nordstr{\"o}m metric to become significant when the field strength at the inner horizon becomes comparable to $b$,
\be 
\frac{Q}{r_-^2} \sim b \, .
\ee
Solving this with $r_- = M - \sqrt{M^2 - Q^2}$ and expanding taking $b$ to be large, we find that the Maxwell approximation can be expected to fail for $Q \lesssim M^{2/3}{b^{-1/3}}$. This is precisely the same scaling as in~\eqref{eq:BI-bound} and hence any NLE will yield the same behaviour, up to overall numerical constants. Moreover, provided that $Q \gg M^{2/3} \ell_P^{1/3}$ the curvature at the inner horizon is sub-Planckian, which ensures we can trust that GR does not receive higher-order or quantum corrections. If we consider the Born-Infeld bound on $b$ to be typical, then we have a range (in geometric units) $9.59 \times 10^{-8} \, \, {\rm m}^{-1} \lesssim b \ll  6.19 \times 10^{34} \, \, {\rm m}^{-1}$ for which we can realize the necessary separation of scales.\footnote{Another example of a theory where the computation can be done explicitly is the recently studied RegMax electrodynamics~\cite{Hale:2023dpf}. For example, in RegMax the self-energy also exhibits the $\alpha Q^{3/2}$ behaviour like Born--Infeld -- see eq.~(39) of~\cite{Hale:2023dpf}. Hence, the qualitative behaviour of the bound in that theory would be identical to~\eqref{eq:BI-bound}, consistent with the general argument we have provided.}

\section{Conclusions}
Cauchy horizons are a pathological feature of Maxwell charged black holes in general relativity. We have shown that a simple modification of Maxwell's theory, employing the framework of nonlinear electrodynamics with finite self-energy, removes such pathologies in the weakly charged regime, imposing a new effective bound on the charge to mass ratio of `physically viable' black holes. According to the conjecture in \cite{Tahamtan:2023tci}, such black holes (contrary to their RN-branch cousins) are also likely stable and their perturbations are `well-posed'.\footnote{ Interestingly, demanding the absence of Cauchy horizons entirely (including the strongly charged regime) is not possible within the (pathology free) NLE framework.}

The sensitivity of the Cauchy horizon under deformations of the theory has been observed also in the case of asymptotically anti-de Sitter black holes~\cite{Hartnoll:2020rwq, Cai:2020wrp, DeClerck:2023fax}. In particular, these examples consider Einstein--Maxwell-$\Lambda$ theory coupled to charged scalars or consider deformations of Maxwell's theory that introduce additional massive gauge fields. Taken together with the results here, these observations further solidify the idea that the Reissner--Nordstr{\"o}m Cauchy horizon is finely tuned both in the space of metric deformations \textit{and} in the space of theories. However, by contrast, while the aforementioned mechanisms require additional fields and anti-de Sitter asymptotics, our observations hold true in pure electrodynamics and irrespective of the asymptotics, be them AdS, dS, or flat.

Of course, the essence of our argument is not restricted to electrodynamics but is a consequence of matter theories with finite self-energy. Thus, inspired by this observation, we may consider a class of more general geometries coupled to matter field sources with finite self-energies, in which the central singularity stays well-behaved and space-like when $M>M_m$, for some \emph{marginal mass} $M_m$ comprised of all other forms of energy the black hole may possess. In particular, here  we had $M_m=U_{\mbox{\tiny self}}^{(0)}$.

In our study we have fully focused on charged spherically symmetric spacetimes, and it is natural to wonder as to what would happen in the rotating case. Namely, could a similar mechanism be applied to the more astrophysically relevant case of the Kerr metric, replacing its Cauchy horizon with a spacelike singularity? As the Kerr Cauchy horizon is not related to matter self-energy, it seems any mechanism to eliminate it would necessarily require a modification of gravitational dynamics. Nonetheless, as the `rotational energy' terms in the metric function have a similar structure to those of Maxwell theory,
\be 
\frac{\Delta}{r^2}=1-\frac{2(M-U)}{r}\,,\quad U=\frac{Q^2+a^2}{2r}\,,
\ee
it is tempting to speculate that such a construction would be possible. 
While such metrics can easily be constructed off-shell, e.g. \cite{Leon:2024uao}, it remains to be seen whether they can also be realized as a solution of any properly modified (e.g. limiting curvature) gravity theories.

\subsection*{Acknowledgements}

We thank Pablo Bueno, Pablo Cano, Roberto Emparan, {\'A}ngel Murcia, Eric Poisson, and Simon Ross for helpful comments and reading an earlier draft of this work. D.K. and T.H. are grateful for support from GA{\v C}R 23-07457S grant of the Czech Science Foundation and the Charles University Research Center Grant No. UNCE24/SCI/016.

\appendix

\section{Proof of horizon structure}

If a NLE violates any of the energy conditions, the corresponding theory is acausal~\cite{Russo:2024xnh}. Hence is it important to assess the implications of these inequalities for the structure of the interior. Here we will focus primarily on the strong energy condition, which is satisfied by, for example, both Born-Infeld and RegMax electrodynamics. We wish to address the following potential issue. In the main text we have constrained the small-$r$ asymptotic behaviour for NLEs with finite self energy, showing the generic behaviours are either Schwarzschild-like or RN-like. However, in principle one could imagine `designing' an NLE which has these small-$r$ asymptotics but an in-principle arbitrary number of additional inner horizons, see e.g.~\cite{Gao:2021kvr}. We will prove that this is not possible for causal NLEs by showing that additional inner horizons would generically violate the~SEC.

Using \eqref{eq:solution} and \eqref{eq:self energy} we may write the metric function in terms of the self-energy,
\be 
f(r) = 1 - \frac{2M}{r} + \frac{2U(r)}{r}\,.
\ee
The strong energy condition requires $\left(T_{\mu\nu} - 1/2T g_{\mu\nu} \right)u^{\mu} u^{\nu} \ge 0$ for all timelike curves $u^\mu$. Restricting to radial timelike curves and using the field equations we obtain,
\be 
{\rm SEC} \quad \Rightarrow \quad  U''(r) \ge 0 \, . 
\ee

Now, suppose we have a finite $U_{\tiny \rm self}$ with $U_{\tiny \rm self}^{(0)} < M$ so that we have the Schwarzschild-like behaviour near $r=0$. Then we have $f(r \to 0) \to -\infty$ while $f(r\to \infty) \to 1$ by asymptotic flatness. Define a new function
\be 
g(r) \equiv r f(r) = r - 2M + 2U(r) \, .
\ee
Note that the zeroes of $f(r)$ are also zeroes of $g(r)$.  Moreover, because $U_{\tiny \rm self}^{(0)} < M$ by assumption, $g(r)$ cannot have an additional zero at $r = 0$ but approaches a finite negative value there. Now, since $g''(r) = 2 U''(r)$ the SEC implies that $g''(r) \ge 0$ while the asymptotics of $f(r)$ imply that $g(r) \to \infty$ as $r\to\infty$ and $g(r) \to (U_{\tiny \rm self}^{(0)} - M) < 0$ as $r \to 0$. Suppose that in addition to the event horizon we have one or more inner horizons. Because of the large and small $r$ asymptotics of $g(r)$, an inner horizon generically implies a maximum of $g(r)$ in the vicinity of which we would have $g''(r) < 0$. Essentially, the SEC requires $g(r)$ to be convex which combined with its constraints at $r = 0$ and $r \to \infty$ is incompatible with more than one zero. Hence no such inner horizon can exist. 

Now consider instead the situation with RN-like small-$r$ asymptotics. The argument runs essentially in the same manner except now as $r\to 0$ we have $g(r) \to (U_{\tiny \rm self}^{(0)} - M) > 0$ by assumption. This means that $g(r)$ can have one (possibly degenerate) inner horizon while maintaining consistency with the SEC. Thus, the existence of any additional  horizons beyond one for the S-branch or two for the RN-branch generically constitutes a violation of the strong energy condition and therefore a violation of causality in the NLE. The interior causal structure for physically viable NLEs is fully determined by the discussion in the main text.

%\bibliography{refs.bib}

%

\end{document}